\documentclass[a4paper,11pt]{article}
\pdfoutput=1
\usepackage{jheppub}
\usepackage{xcolor,bm,comment,braket}
\usepackage[T1]{fontenc}
\usepackage{comment}
\newcommand{\del}{\partial}
\newcommand{\beq}{\begin{eqnarray}}
\newcommand{\eeq}{\end{eqnarray}}

\newcommand{\tr}{\mathop{\mathrm{tr}}}
\newcommand{\SU}{\text{SU}}
\newcommand{\U}{\text{U}}
\newcommand{\rmi}{\text{i}}
\newcommand{\rme}{\text{e}}

\begin{document}

\title{
Solitonic ground state in supersymmetric theory in background 
}

\author[a,b]{Muneto Nitta}
\emailAdd{nitta(at)phys-h.keio.ac.jp}

\author[c]{and Shin Sasaki}
\emailAdd{shin-s(at)kitasato-u.ac.jp}

\affiliation[a]{Department of Physics \& 
Research and Education Center for Natural Sciences,
Keio University, 4-1-1 Hiyoshi, Kanagawa 223-8521, Japan}
\affiliation[b]{
International Institute for Sustainability with Knotted Chiral Meta Matter(SKCM$^2$), Hiroshima University, 1-3-2 Kagamiyama, Higashi-Hiroshima, Hiroshima 739-8511, Japan
}
\affiliation[c]{
Department of Physics,  Kitasato University, 
  Sagamihara 252-0373, Japan
}

\abstract{
A solitonic ground state called a chiral soliton lattice (CSL) 
is realized in a supersymmetric theory with 
background magnetic field and finite chemical potential. 
To this end, we construct, in the superfield formalism,  
a supersymmetric chiral sine-Gordon model as a neutral pion sector of a supersymmetric two-flavor chiral Lagrangian with a Wess-Zumino-Witten term. 
The CSL ground state appears in the presence of 
either a strong magnetic field 
and/or large chemical potential, or 
a background fermionic condensate  
in the form of  a fermion bilinear consisting of 
the gaugino and a superpartner of a baryon gauge field.
}

\maketitle

\section{Introduction}

Determination of 
the ground states or vacua and phase structures 
is quite important to 
understand any physical system.
For such a purpose, 
it is becoming important to consider a possibility of
spatially inhomogeneous ground states. 
Several examples can be found 
in condensed matter physics such as superconductors \cite{Fulde:1964zz,Larkin:1964wok,Machida:1984zz,Yoshii:2011yt} 
and polyacetylene \cite{Su:1979wc,Takayama:1980zz,Niemi:1984vz}, 
and in quantum field theories such as 
the Gross--Neveu and Nambu--Jona-Lasino models~\cite{Basar:2008im,Basar:2008ki,Basar:2009fg} 
and quantum chromodynamics (QCD)  \cite{Nakano:2004cd,Nickel:2009ke,Casalbuoni:2003wh,Anglani:2013gfu,Buballa:2014tba}.
For such a kind of ground states, the order parameter is characterized by a spatially varying function 
and several translational symmetries are spontaneously broken there.
However, without any external field, spatial modulations are unstable due to the so-called Landau-Peierls instability \cite{Hidaka:2015xza,Lee:2015bva}.

On the other hand, solitonic ground states are present in 
the presence of spin-orbit coupling in condensed matter systems or 
background fields in quantum field theories such as a magnetic field or rotation. 
The examples of the former contain chiral magnets 
and ultracold atomic gases with synthetic gauge fields.
The ground states of the chiral magnets contain two kinds of  inhomogeneous ground states made of 
topological solitons 
due to the presence of the 
so-called Dzyaloshinskii-Moriya (DM) interaction \cite{Dzyaloshinskii,Moriya:1960zz}.
One is a chiral soliton lattice (CSL), 
also called a spiral phase \cite{togawa2012chiral,KISHINE20151,PhysRevB.97.184303,Ross:2020orc,Amari:2023gqv}, 
where the energy of a single soliton 
or domain wall is negative 
and thus one dimensionally inhomogeneous states 
composed of solitons or domain walls have a lower energy than 
uniform states. 
The other is the 2D Skyrmion lattice (crystal) phase where 
the energy of a single Skyrmion is negative \cite{doi:10.1126/science.1166767,doi:10.1038/nature09124,doi:10.1038/nphys2045,Rossler:2006,Lin:2014ada,Han:2010by,Ross:2020hsw}. 
Ultracold atomic gases with spin-orbit couplings 
or synthetic gauge fields \cite{Lin:2009vfu,Goldman:2013xka}
give the solitonic ground states of 
2D Skyrmions \cite{PhysRevA.84.011607} 
and
3D Skyrmions
\cite{Kawakami:2012zw}.

As the example of the latter, 
QCD under extreme conditions like high baryon density, pronounced magnetic fields, and rapid rotation has been extensively studied.
Recently, the QCD phase diagram including such conditions 
gathers significant attention 
due to the relevance in the interior of neutron stars 
and heavy-ion collisions 
\cite{Fukushima:2010bq}. 
In particular, 
QCD in strong magnetic fields has received quite intense attention.
At low energy, QCD can be described 
in terms of pions degrees of freedom 
by the chiral Lagrangian 
or chiral perturbation theory (ChPT) 
 accompanied with 
the Wess-Zumiono-Witten (WZW) term 
\cite{Leutwyler:1993iq}.
In the presence of a background magnetic field $B$
at finite chemical potential $\mu_{\textrm{B}}$,
the WZW term in QCD with two flavors
contains 
an anomalous coupling of the neutral pion $\pi_0$ to the magnetic field via the chiral anomaly 
\cite{Son:2004tq,Son:2007ny}
through the Goldstone-Wilczek current \cite{Goldstone:1981kk,Witten:1983tw}.
Then, 
when the background gauge field/and or 
chemical potential are large enough
\begin{equation}
    B  \mu_{\textrm{B}}
    \geq 
    \frac{16\pi m_{\pi}}{f_{\pi}^2 e }, 
    \label{eq:critical}
\end{equation}
    with the pion's mass $m_{\pi}$ and decay constant $f_{\pi}$ and the gauge coupling constant $e$,
the ground state 
 is inhomogeneous in the form of a CSL 
consisting of a stack of solitons 
\cite{Son:2007ny,Eto:2012qd,Brauner:2016pko},
analogous to chiral magnets. 
To show this, neglecting  
charged pions degrees of freedom,  
the neutral pion sector of 
the chiral Lagrangian reduces 
to the so-called chiral sine-Gordon model, in common with chiral magnets \cite{togawa2012chiral,KISHINE20151,PhysRevB.97.184303}. 
It was also discussed that thermal fluctuations enhance the stability of CSL~\cite{Brauner:2017uiu,Brauner:2017mui,Brauner:2021sci,Brauner:2023ort}.
Similar CSLs of the $\eta$ 
(or $\eta'$) meson also appear 
under rapid rotation instead of strong magnetic field~\cite{Huang:2017pqe,Nishimura:2020odq,Chen:2021aiq,Eto:2021gyy,Eto:2023tuu,Eto:2023rzd}.
Further investigations have been done 
into the quantum nucleation of CSLs \cite{Eto:2022lhu,Higaki:2022gnw},  
quasicrystals \cite{Qiu:2023guy}, 
the domain-wall Skyrmion phase \cite{Eto:2023lyo,Eto:2023wul},
the interplay with Skyrmion crystals at zero magnetic field  \cite{Kawaguchi:2018fpi,Chen:2021vou,Chen:2023jbq},
and an Abrikosov's vortex lattice and baryon crystals~\cite{Evans:2022hwr,Evans:2023hms}.
 Among various studies of CSL, 
one of the most important directions 
is the stability of CSL 
beyond perturbations;
CSL states in QCD-like theory such as 
$\SU(2)$ QCD and vector-like gauge theories were studied in 
Refs.~\cite{Brauner:2019rjg,Brauner:2019aid} 
for nonperturbative studies in lattice gauge theories 
without the sign problem.
On the other hand, the motivation of this paper is to investigate a CSL in 
supersymmetric theories which should be 
also relevant for studies of its nonperturbative aspects.

Supersymmetry (SUSY) is 
a symmetry between bosons and fermions 
and was quite extensively studied for long time 
because of a lot of reasons 
\cite{Nilles:1983ge}: it was expected to solve the so-called naturalness problem in the phenomenological side \cite{Haber:1984rc,Kane:1993td},
while in the theoretical side it is a useful or necessary tool 
to control nonperturbative effects 
in quantum field theory and string theory 
 \cite{Seiberg:1994aj,Seiberg:1994rs}. 
Bogomol'nyi-Prasad-Sommerfiled (BPS)  topological solitons and instantons 
preserve some fraction of SUSY and 
are nonperturbatively stable
\cite{Tong:2005un,Eto:2006pg,Shifman:2007ce,Shifman:2009zz,Tong:2008qd}, 
thereby playing a crucial role for study of nonperturbative aspects.  
However, in reality SUSY is not present at low energy in nature, and 
thus SUSY breaking is one of the most important aspects in phenomenology \cite{Intriligator:2006dd}. 
Among various SUSY breaking mechanisms, 
a unconventional SUSY breaking mechanism  
due to a spatial or temporal modulation was  proposed \cite{Nitta:2017yuf,BjarkeGudnason:2018aij} 
by extending  modulated vacua in 
relativistic bosonic field theories 
\cite{Nitta:2017mgk,Gudnason:2018bqb}.
To this end, SUSY higher derivative terms free from 
ghost and auxiliary field problem  \cite{Khoury:2010gb,Khoury:2011da,Koehn:2012ar,Adam:2011hj,Farakos:2012qu,Nitta:2014pwa,Nitta:2014fca,Nitta:2015uba,Nitta:2020gam} were essential.
By contrast, in the configuration proposed in this paper, SUSY is broken 
by a solitonic ground state (as well as background fields).
It is interesting to investigate how the effects of background fields
 break SUSY and modifies the well-known relations of states and energies
 in SUSY theories.
In SUSY theories, the relations among vacua, BPS states and their
 energies are rigorously determined.
For example, the energies of states in SUSY theories are positive semi-definite in general
and the vacuum energy is exactly zero. 
We also know that the energy of a BPS soliton state is completely
 determined by the boundary conditions, which is
 always greater than the vacuum energy. 
We will see that, even within the framework of SUSY theory, the
 effects of the background fields drastically change these relations of
 energies of the BPS and vacuum states.

In this paper, 
we discuss, in the superfield formalism,  
a manifestly supersymmetric  
extension of $\SU(2)$ 
chiral Lagrangian with the WZW term 
in magnetic field at finite density. 
Instead of the full Lagrangian, 
we succeed to construct the neutral pion sector 
which is a SUSY chiral sine-Gordon model. 
We then construct 
a BPS domain wall (chiral soliton) 
and show that in a certain parameter region of strong magnetic field 
and/or large chemical potential, 
the tension of the domain wall becomes negative.
In such a case, 
the ground state is CSL, an alternate array of 
BPS and anti-BPS chiral solitons, 
where SUSY is broken.
While this is the same with conventional QCD, 
we also find that a CSL also occurs in a constant fermion condensation background.

Some comments on previous works are in order. 
First,
a SUSY sine-Gordon model was discussed 
in the literature \cite{DiVecchia:1977nxl,Hruby:1977nc,Marinucci:1979ke,Cassandro:1989bb,Cassandro:1989vy}, but 
SUSY {\it chiral} sine-Gordon model was not.
Second, 
the WZW term has the same form with the previously known 
supersymmetric WZW term \cite{Nemeschansky:1984cd,Nitta:2000ebt}.\footnote{
See  Refs.~\cite{Gates:1995fx,Gates:1996cq,Gates:2000rp} 
for further studies of SUSY WZW terms, 
see also Refs.~\cite{Buchbinder:1994xq,Buchbinder:1994iw,Banin:2006db}.
}
Third,
a BPS domain wall in SUSY chiral Lagrangian was studied 
without the WZW term \cite{Gudnason:2016frn}.

This paper is organized as follows.
In Sec.~\ref{sec:CSL}, we summarize ChPT in QCD in the magnetic field 
at finite density and 
the CSL ground state in it. 
In Sec.~\ref{sec:susy-CPT}, 
we construct manifestly supersymmetric 
chiral Lagrangian and WZW term 
in the superfield formalism.
In Sec.~\ref{sec:soliton}, 
we construct BPS solitons (domain walls), 
and 
discuss solitonic ground states (CSL)
consisting of an array of BPS and anti-BPS solitons 
in the presence of an external gauge field background or 
a fermion background.
Section \ref{sec:summary} is devoted to a summary and discussion.
In Appendix \ref{sec:FI}, 
we show the incompatibility of
the Fayet-Iliopoulos term.

%\newpage

%%%%%%%%%%%%%%%%
\section{Chiral soliton lattice in strong magnetic field: a review} \label{sec:CSL}

In this section, we give a brief review on 
the two-flavor chiral Lagrangian with the WZW term 
at finite density and background gauge field, 
and the CSL ground state.

%%%%%%
\subsection{Chiral Lagrangian}
We concentrate on the phase in which the chiral symmetry is
spontaneously broken.
The low-energy dynamics in this phase is described by the two-flavor ChPT, which is an effective field theory for pions.
The pion fields $\phi_a \, (a=1,2,3)$ are represented by a $2\times 2$ unitary matrix
$\Sigma = \rme^{\rmi \tau_a \frac{\phi^a}{f_{\pi}}}$ which undergoes the
$\SU(2)_{\mathrm{L}} \times \SU(2)_{\mathrm{R}}$ chiral symmetry;
\begin{equation}
\Sigma \rightarrow L \Sigma R^{\dagger}.
\end{equation}
where both $L$ and $R$ are $2 \times 2$ unitary matrices.
Here $\tau_a \, (a = 1, 2, 3)$ are the Pauli matrices normalized as
$\tr[\tau_a \tau_b] = 2 \delta_{ab}$.

The $\mathrm{U(1)}_{\mathrm{em}}$ electromagnetic gauge transformation
is given by
\begin{equation}
\Sigma \to \rme^{\mathrm{i} f_{\pi}^{-1} \lambda\frac{\tau_3}{2}} \Sigma
 \rme^{-\mathrm{i} f_{\pi}^{-1} \lambda\frac{\tau_3}{2}} 
\quad \text{and} \quad 
A_m \to A_m - \frac{1}{e} \partial_m \lambda,
\end{equation}
where $\lambda$, $e$ are the gauge parameter and the coupling constant.
The associated covariant derivative $D_{m}$ is defined by
\begin{gather}
    D_{m} \Sigma
    = \del_{m} \Sigma
    + \rmi e A_{m} [Q, \Sigma] \label{def_cov_del} \,, \\
    Q = \frac{1}{6}\bm{1}_2 + \frac{1}{2}\tau_3
    \,,
\end{gather}
where the matrix $Q$ represents the electric charge carried by quarks.
The effective Lagrangian at ${\cal O}(p^2)$ is given by
\begin{equation}
    \mathcal{L} = \mathcal{L}_{\mathrm{ChPT}} + \mathcal{L}_{\mathrm{WZW}} 
    \label{eq:total-L}
\end{equation}
where the first term is given by
\begin{gather}
    \mathcal{L}_{\textrm{ChPT}}
    = - \frac{f_{\pi}^2}{4} \tr \left(D_{m}\Sigma D^{m}\Sigma^{\dag} \right) 
    - \frac{f_{\pi}^2m_{\pi}^2}{4}
    \tr 
    (2 {\bf 1}_2 -\Sigma-\Sigma^{\dag}).
\label{ChPT_with_B}
\end{gather}
The second term in \eqref{eq:total-L} gives the coupling of the pions to the external
$\U(1)_{\text{B}}$ baryon gauge field $A_m^{\mathrm{B}} = (\mu_{\rm B},0,0,0)$ through the WZW term \cite{Son:2007ny}.
This is given in Refs.~\cite{Son:2004tq,Son:2007ny} as
\begin{equation}
\mathcal{L}_{\mathrm{WZW}} = - \left( A^{\mathrm{B}}_{m} + \frac{e}{2} A_m \right) j_{\mathrm{GW}}^{m}.
\label{effective_Lagrangian_GW}
\end{equation}
Here, $j_{\mathrm{GW}}^m$ is the Goldstone-Wilczek current
\cite{Goldstone:1981kk,Witten:1983tw} defined by
\begin{equation}
j_{\mathrm{GW}}^{m} 
= - \frac{\epsilon^{mnpq}}{24\pi^2} \mathrm{tr} 
\left( 
L_{n} L_{p} L_{q} - 3 \mathrm{i} e \partial_{n} 
\left[ 
A_{p} Q (L_{q} + R_{q})
\right] 
\right),
\label{eq:GW_current}
\end{equation}
where $L_m = \Sigma \del_m \Sigma^{\dagger}$, $R_m = \del_m
\Sigma^{\dagger} \Sigma$ are the left-invariant and the right-invariant
Maurer-Cartan 1-forms respectively.

It is convenient to rewrite the Goldstone-Wilczek current \eqref{eq:GW_current} in the
form where the $\U(1)_{\text{em}}$ gauge symmetry is manifest \cite{Son:2007ny};
\begin{align}
j^{m}_{\text{GW}} = - \frac{1}{24 \pi^2} \varepsilon^{mnpq} \mathrm{tr}
\Bigg[
(\Sigma D_{n} \Sigma^{\dagger}) (\Sigma D_{p} \Sigma^{\dagger})
 (\Sigma D_{q} \Sigma^{\dagger}) 
- \frac{3 e i}{2} F_{np} Q (\Sigma D_{q} \Sigma^{\dagger}
 + D_{q} \Sigma^{\dagger} \Sigma)
\Bigg].
\label{eq:GW_current_2}
\end{align}
In the low energies, only the neutral pion $\phi^3 = \phi$ contributes to physics.
We therefore set $\Sigma = e^{i \tau_3 \frac{\phi}{f_{\pi}}}$ and 
the covariant derivative $D_{m}$ is replaced by $\del_{m}$
\begin{align}
D_{m} \Sigma = \del_{m} \Sigma + i e A_{m} \left[\frac{\tau_3}{2},
 \Sigma \right] = \del_{m} \Sigma.
\end{align}
Then the ChPT Lagrangian for the neutral pion is given by
\begin{align}
\mathcal{L}_{\mathrm{ChPT}} =& \ 
- \frac{f_{\pi}^2}{4} 
\mathrm{tr} 
\Big(
\del_m \Sigma^{\dagger} \del^m \Sigma
\Big)
+ 
\frac{f_{\pi}^2 m^2_{\pi}}{4}
\Big(
\mathrm{tr} \Sigma
+ 
\text{h.c.}
\Big)
- \frac{f_{\pi}^2 m_{\pi}^2}{2} \mathrm{tr} \mathbf{1}_2
\notag \\
& \ 
+
\frac{1}{24 \pi^2} 
\varepsilon^{mnpq} 
\left(
A_m^B + \frac{e}{2} A_m
\right)
\mathrm{tr}
\Bigg[
(\Sigma \del_n \Sigma^{\dagger})
(\Sigma \del_p \Sigma^{\dagger})
(\Sigma \del_q \Sigma^{\dagger})
-
\frac{3i}{2} F_{np} \tau_3 \Sigma \del_q \Sigma^{\dagger}
\Bigg].
\label{eq:ChPT_pi0}
\end{align}

%%%%%%%%%%%%%%%%%%%%
\subsection{Chiral soliton lattice}
It is noteworthy to see that the model for the neutral pion
\eqref{eq:ChPT_pi0} is given by the sine-Gordon model;
\begin{eqnarray}
\mathcal{L}
= 
- \frac{1}{2}(\del_{m} \phi)^2
- f_{\pi}^2m_{\pi}^2
\Bigg\{
1 - \cos \left(\frac{\phi}{f_{\pi}} \right)
\Bigg\}
+ 
\frac{e\mu_{\textrm{B}}}{4\pi} \vec{B} \cdot \vec{\nabla} \phi,
\label{eq:SG}
\end{eqnarray}
where we have assumed the external ${\mathrm U}(1)_{\text{em}}$ constant magnetic
field $\vec{B}$ and that the pion field is time independent $\del_0 \Sigma = 0$. 
This model is known as the chiral sine-Gordon model 
studied extensively in the context of chiral magnets 
\cite{togawa2012chiral,KISHINE20151,PhysRevB.97.184303,Ross:2020orc,Amari:2023gqv}.

The QCD background in the absence of the external background field is
given by  $\phi/f_\pi = 2\pi n$ with 
$n \in {\mathbb Z}$.
where the energy (density) vanishes. 
The last term in \eqref{eq:SG} does not contribute to the
equation motion:
\begin{align}
\nabla^2 \phi - f_{\pi} m_{\pi}^2 \sin \frac{\phi}{f_{\pi}} = 0.
\end{align}
The simplest solution to this equation is known as 
a single sine-Gordon (anti-)kink 
interpolating two adjacent vacua: 
\begin{align}
 \phi (z) = 4 f_{\pi} \mathrm{arctan} \big[ e^{\pm m_{\pi} (z - c)} \big],
\label{eq:sg_kink}
\end{align}
where $c$ is the position of the center or the translational modulus.
The tension (energy per unit area) is given by
\begin{align}
E
=
& \ \int_{- \infty}^{\infty} \! dz \, 
\Bigg[
\frac{1}{2} (\del_z \phi)^2 
+ f_{\pi}^2 m_{\pi}^2 
\Big(
 1 - \cos \left( \frac{\phi}{f_{\pi}} \right) 
\Big)
- \frac{e f_{\pi} \mu_{\rm B} B_z}{4 \pi^2} \del_z \phi
\Bigg]
\notag \\
=& \ 8 m_{\pi} f_{\pi}^2 
\mp \frac{e \mu_{\rm B} B_z}{2 \pi}. \label{eq:tension}
\end{align}
It is obvious that the finite baryon density and the external magnetic field give 
a negative (positive) contribution to the energy 
for a sine-Gordon (anti-)kink. 
This imbalance between 
a kink and an anti-kink 
is the origin of the term ``{\it chiral}''
of the chiral sine-Gordon model.
At the boundary of Eq.~(\ref{eq:critical}),  
the tension of a single sine-Gordon kink 
in Eq.~(\ref{eq:tension}) 
with the upper sign 
becomes zero, 
$E=0$, 
and the ground state is 
a single kink degenerated with the QCD vacuum.
For larger background satisfying 
inequality of Eq.~(\ref{eq:critical}), 
one finds that the tension of the sine-Gordon kink is negative, $E<0$, and the kink in Eq.~\eqref{eq:sg_kink} with the upper sign 
is energetically more
stable against the ordinary QCD vacuum $\phi/f_\pi = 2 n \pi$. 
However, one cannot create infinite numbers of kinks since two adjacent sine-Gordon kinks repel each other, 
increasing the energy. 
Consequently, the ground state in the presence of the large background fields is 
a stack of sine-Gordon kinks, 
that is a CSL \cite{Son:2007ny,Eto:2012qd,Brauner:2016pko}.

\section{Supersymmetric chiral perturbation theory}\label{sec:susy-CPT}
In this section, we construct a four-dimensional $\mathcal{N} = 1$
supersymmetric ChPT in the external background gauge fields.
The basic elements are the supersymmetric generalization of the pion
field $\Sigma$ and the $\mathrm{U}(1)_{\text{em}}$ and the baryon gauge
fields $(A_m, A^B_m)$.
In the following, we follow the Wess-Bagger conventions \cite{Wess:1992cp} of superfields.

First, the target space of SUSY nonlinear sigma models must be K\"{a}hler, 
and thus pion fields $\mathrm{SU}(2)$ 
must be complexified:
\begin{eqnarray}
 \mathrm{SU}(2)^{\mathbb{C}} \simeq 
 \mathrm{SL}(2,{\mathbb C}) \simeq 
 T^*  \mathrm{SU}(2).
\end{eqnarray}
We then introduce the $\mathrm{SU}(2)^{\mathbb{C}}$-valued chiral superfield $\mathbf{\Sigma}$ whose lowest
component is given by $\Sigma$;
\begin{align}
\mathbf{\Sigma} (y, \theta) = \Sigma + \sqrt{2} \psi_{\Sigma} \theta +
 F_{\Sigma} \theta^2 \in \mathrm{SU}(2)^{\mathbb{C}}
\end{align}
where $\psi_{\Sigma}, F_{\Sigma}$ are the Weyl fermion and the auxiliary
field and $(y,\theta)$ are the chiral coordinates.
The real parts of $\Sigma$ denote Nambu-Goldstone 
(NG) bosons (pions) associated with the chiral symmetry breaking while the imaginary parts of $\Sigma$ are 
so-called quasi-NG bosons which are not associated with a symmetry breaking but are required from SUSY \cite{Bando:1983ab,Bando:1984cc,Bando:1984fn,Bando:1987br,Shore:1988mn,Kotcheff:1988ji,Higashijima:1997ph,Nitta:1998qp}.
On the other hand, $\psi_{\Sigma}$ are called quasi-NG fermions. 
The kinetic and the mass terms of $\mathbf{\Sigma}$ are given by\footnote{
More generally, the K\"{a}hler potential can be 
an arbitrary function $f$ of two variables:
$\int \! d^4 \theta \, f \left(
\mathrm{tr} 
\big[
\mathbf{\Sigma}^{\dagger} \mathbf{\Sigma}
\big]), 
\mathrm{tr} 
\big[
\mathbf{\Sigma}^{\dagger} \mathbf{\Sigma}
\big]^2 \right)
$. This corresponds to the degrees of freedom 
to deform noncompact directions corresponding to 
quasi-NG bosons, which cannot be fixed by the 
real symmetry $\mathrm{SU}(2)$ \cite{Shore:1988mn,Kotcheff:1988ji,Nitta:1998qp}.
}
\begin{align}
\mathcal{L}_{\text{kin}} = \frac{2}{\beta^2} \int \! d^4 \theta \,
 \mathrm{tr} 
\big[
\mathbf{\Sigma}^{\dagger} \mathbf{\Sigma}
\big]
+ 
\left(
\frac{2m}{\beta^2} \int \! d^2 \theta \, \mathrm{tr} \mathbf{\Sigma} +
 \text{h.c.}
\right).
\end{align}
Here the trace is taken over the $2 \times 2$, SU(2)$^{\mathbb{C}}$-valued matrices and 
$m,\beta$ are real parameters of dimensions $+1$ and $-1$, respectively.
Later, we will see that 
the correspondences of these parameters with those in the last section are 
$1/\beta = f_{\pi}$ and $m = m_{\pi}^2$.

The WZW part of the coupling to the external field $A^B_m$ in the Goldstone-Wilczek
current is constructed with the help of the SUSY WZW term
elucidated in Refs.~\cite{Nemeschansky:1984cd,Nitta:2000ebt};
\begin{align}
\mathcal{L}_{\text{WZW}} = \int \! d^4 \theta \, 
V^B 
\mathrm{tr}
\Big[
(\mathbf{\Sigma} (\sigma^m)_{\alpha \dot{\alpha}} \del_m \mathbf{\Sigma}^{\dagger})
(\mathbf{\Sigma} \bar{D}^{\dot{\alpha}} \mathbf{\Sigma}^{\dagger})
(D^{\alpha} \mathbf{\Sigma} \mathbf{\Sigma}^{\dagger})
\Big] + \text{h.c.}
\label{eq:SUSY_WZW}
\end{align}
Here $V^B$ is the ${\mathrm U}(1)$ vector multiplet for the external baryon number
gauge field $A^B_m$. 
This is given, in the Wess-Zumino gauge, by
\begin{align}
V^B (x, \theta, \bar{\theta}) = 
- (\theta \sigma^m \bar{\theta}) A^B_m + i \theta^2 \bar{\theta}
 \bar{\lambda}^B - i \bar{\theta}^2 \theta \lambda^B + \frac{1}{2}
 \theta^2 \bar{\theta}^2 D^B,
\end{align}
where $\lambda^B, D^B$ are the Weyl fermion and the auxiliary field, respectively. 
The last part in the Goldstone-Wilczek current may be given by
\begin{align}
\mathcal{L}_{\text{GW}} = \frac{i}{2} \int \! d^4 \theta \, V^B
 W_{\alpha} \, \mathrm{tr} 
\Big[
\tau_3 \mathbf{\Sigma}^{\dagger} D^{\alpha} \mathbf{\Sigma}
\Big] + \text{h.c.}
\label{eq:SUSY_GW}
\end{align}
where $W_{\alpha} = - \frac{1}{4} \bar{D}^2 D_{\alpha} V$ is the field
strength for the ${\mathrm U}(1)_{\text{em}}$ vector multiplet $V$.
Although Eqs.~\eqref{eq:SUSY_WZW} and \eqref{eq:SUSY_GW} contain derivatives
of superfields, which sometimes give rise to a potential auxiliary
field problem \cite{Gates:1995fx,Gates:1996cq,Gates:2000rp}, they cause no troubles.
We will see this in due course.

In order to see that the above interactions indeed provide a natural supersymmetric
generalization of ChPT, we now write down the component expression of
the Lagrangian $\mathcal{L}_{\text{SChPT}} = \mathcal{L}_{\text{kin}} +
\mathcal{L}_{\text{WZW}} + \mathcal{L}_{\text{GW}}$.
After some calculations, the bosonic part of the Lagrangian is found to be
\begin{align}
\mathcal{L}_{\text{SChPT}} =& \ 
\frac{2}{\beta^2} 
\mathrm{tr}
\Big[
- \del_m \Sigma^{\dagger} \del^m \Sigma
+ 
F_{\Sigma} \bar{F}_{\Sigma}
\Big]
+
\frac{2m}{\beta^2}
\mathrm{tr}
\Big[
F_{\Sigma} + \bar{F}_{\Sigma}
\big]
\notag \\
& \ 
- 2 
\mathrm{tr}
\Big[
\Big(
- \eta^{mn} \eta^{pq} + \eta^{mp} \eta^{nq} + \eta^{mq} \eta^{np} 
+ i 
\varepsilon^{mnpq}
\Big)
A_m^B 
(\Sigma \del_n \Sigma^{\dagger})
(\Sigma \del_p \Sigma^{\dagger})
(\del_q \Sigma \cdot \Sigma^{\dagger})
\notag \\
& \qquad \qquad
+ A_m 
(\Sigma \del^m \Sigma^{\dagger})
\Sigma \bar{F}_{\Sigma} F_{\Sigma} \Sigma^{\dagger} 
\Big]
+ \text{h.c.}
\notag \\
& \ 
+
\frac{i}{2}
\Big(
- A^B_m D \,
\mathrm{tr} 
\big[
\tau_3 \Sigma^{\dagger} \del^m \Sigma
\big]
- A_m^B F^{mn} 
\mathrm{tr} 
\big[
\tau_3 \Sigma^{\dagger} \del_n \Sigma
\big]
\notag \\
& \qquad \qquad \qquad \qquad \qquad \qquad 
- \frac{i}{2} 
\varepsilon^{mnpq} A^B_m F_{pq} 
\mathrm{tr}
\big[
\tau_3 \Sigma^{\dagger} \del_n \Sigma
\big]
\Big)
+ \text{h.c.}
\end{align}
We find that this precisely contains the interaction terms in \eqref{eq:ChPT_pi0}
and is a natural SUSY generalization of the ChPT in the external background
gauge fields.

In order to find the interactions for the neutral pion, we now write 
\begin{align}
\mathbf{\Sigma} = e^{i \frac{\tau^3}{2} \beta \Phi}
= \mathbf{1}_2 \cos \left( \frac{\beta \Phi}{2} \right) + i \tau_3 \sin
 \left( \frac{\beta \Phi}{2} \right)
\end{align}
where $\Phi (y,\theta) = \varphi + \sqrt{2} \theta \psi + \theta^2 F$ is the chiral
superfield for the neutral pion $\phi$.
The real part of the complex scalar field $\varphi$ corresponds to the neutral pion $\phi$ 
(up to the normalization) while the imaginary part is the quasi-NG boson.
Then the superfield Lagrangian becomes
\begin{align}
\mathcal{L}_{\text{SChPT}} =& \ 
\frac{4}{\beta^2} \int \! d^4 \theta \, \cos \left( \frac{\beta}{2}
 (\Phi - \Phi^{\dagger})\right)
+ 
\Bigg[
\frac{4 m}{\beta^2} \int \! d^2 \theta \, \cos \left( \frac{\beta}{2}
 \Phi \right) + \text{h.c.}
\Bigg]
\notag \\
& \
+ 
\Bigg[
\frac{\beta^3}{4} \int \! d^4 \theta \,
V^B \sin \left( \frac{3}{2} \beta (\Phi - \Phi^{\dagger}) \right)
(\sigma^m)_{\alpha \dot{\alpha}} \del_m \Phi^{\dagger}
 \bar{D}^{\dot{\alpha}} \Phi^{\dagger} D^{\alpha} \Phi
+ \text{h.c.}
\Bigg]
\notag \\
& \ 
+ 
\Bigg[
\frac{i \beta}{2} \int \! d^4 \theta \, 
V^B W_{\alpha} \cos \left( \frac{\beta}{2} (\Phi -
 \Phi^{\dagger}) \right) D^{\alpha} \Phi 
+ \text{h.c.}
\Bigg].
\end{align}
It is straightforward to calculate the bosonic part of the Lagrangian.
The result is 
\begin{align}
\mathcal{L}_{\text{SChPT}} =& \ 
- \frac{1}{2 \beta} 
\sin
\left(
\frac{\beta}{2} (\varphi - \bar{\varphi})
\right)
\Box (\varphi - \bar{\varphi})
-
\frac{1}{4}
\cos 
\left(
\frac{\beta}{2} (\varphi - \bar{\varphi})
\right)
\del_m (\varphi + \bar{\varphi}) \del^m (\varphi + \bar{\varphi})
\notag \\
& \ 
+ 
\cos 
\left(
\frac{\beta}{2} (\varphi - \bar{\varphi})
\right)
F \bar{F}
- 
\frac{2 m}{\beta} 
\Bigg\{
 F \sin \left( \frac{\beta}{2} \varphi \right)
+ \bar{F} \sin \left( \frac{\beta}{2} \bar{\varphi} \right)
\Bigg\}
\notag \\
& \
- \frac{\beta^2}{4} \sin \left( \frac{3}{2} \beta (\varphi -
 \bar{\varphi}) \right)
\Big[
(A_m^B \del^m \varphi) (\del_n \bar{\varphi} \del^n \bar{\varphi})
-
(A_m \del^m \bar{\varphi}) (\del_n \varphi \del^n \varphi)
\Big]
\notag \\
& \ 
+  \frac{\beta^2}{2} i F \bar{F} \sin \left( \frac{3}{2} \beta (\varphi -
 \bar{\varphi}) \right) A_m^B \del^m (\varphi - \bar{\varphi})
\notag \\
& \ 
+
\frac{\beta}{2} \cos \left( \frac{\beta}{2} (\varphi - \bar{\varphi}) \right)
\Bigg[
i A^B_m D \, \del^m (\varphi - \bar{\varphi})
- i A^B_m F^{mn} \del_n (\varphi - \bar{\varphi})
\notag \\
& \qquad \qquad \qquad \qquad \qquad \qquad 
+ \frac{1}{2} \varepsilon^{mnpq} A^B_p F_{mn} \del_q (\varphi
 + \bar{\varphi})
\Bigg].
\label{eq:SChPT_superfield2}
\end{align}
We stress that even though the superfield Lagrangian involves derivative
interactions, the auxiliary fields do not have spacetime derivative terms
in the component Lagrangian.
This fact is necessary in the sense that the equations of motion for the
auxiliary fields are algebraic ones and we can write down the explicit interaction terms for $\varphi$.
Indeed, the equation of motion for $\bar{F}$ is solved by
\begin{align}
F = \frac{2 m}{\beta} \sin \left( \frac{\beta}{2} \bar{\varphi} \right)
\left\{
\cos \left( \frac{\beta}{2} (\varphi - \bar{\varphi})\right) 
+ \frac{\beta^2}{2} i
\sin \left( 
\frac{3}{2} \beta (\varphi - \bar{\varphi}) 
\right) A_m^B \del^m (\varphi - \bar{\varphi})
\right\}^{-1}.
\end{align}
After integrating out the auxiliary fields $F, \bar{F}$, we find that the
potential term together with the derivative interactions appear in the Lagrangian:
\begin{align}
\mathcal{L}_{\text{SChPT}} =& \ 
\frac{1}{\beta} \sinh (\beta \chi) \Box \chi
-
\cosh (\beta \chi) \del_m \phi \del^m \phi
\notag \\
& \ 
- \frac{4m^2}{\beta^2}
\Bigg\{
\cos^2 \left( \frac{\beta \phi}{2} \right) \sinh^2 \left( \frac{\beta
 \chi}{2} \right)
+ \sin^2 \left( \frac{\beta \phi}{2} \right) \cosh^2 \left( \frac{\beta
 \chi}{2} \right)
\Bigg\}
\times 
\notag \\
&  \qquad \qquad \qquad \times
\Bigg\{
\cosh (\beta \chi) - \beta^2 \sinh (3 \beta \chi) A_m^B \del^m \chi
\Bigg\}^{-1}
\notag \\
& \ 
+ \frac{\beta^2}{2} \sinh (3 \beta \chi)
\Bigg[
- 2 (A_m^B \del^m \phi) (\del_n \phi \del^n \chi)
+ (A_m^B \del^m \chi) (\del_n \phi \del^n \phi - \del_n \chi \del^n \chi )
\Bigg]
\notag \\
& \ 
+ \beta \cosh (\beta \chi) 
\Bigg[
- D \, A_m^B \del^m \chi 
+ A_m^B F^{mn} \del_n \chi
+ \frac{1}{2} \varepsilon^{mnpq} A_p^B F_{mn} \del_q \phi
\Bigg].
\label{eq:SChPT_component}
\end{align}
Here, for later convenience, we have decomposed the scalar field into 
the real and the imaginary parts $\varphi = \phi + i \chi$.

Lorentz invariant vacua are found for the constant scalar fields $\phi$ and $\chi$.
They are specified by the zeros of the potential
\begin{align}
V = \frac{4m^2}{\beta^2} 
\Bigg\{
\cos^2 \left( \frac{\beta \phi}{2} \right) \sinh^2 \left( \frac{\beta
 \chi}{2} \right)
+ \sin^2 \left( \frac{\beta \phi}{2} \right) \cosh^2 \left( \frac{\beta
 \chi}{2} \right)
\Bigg\} 
\Big\{
\cosh(\beta \chi)
\Big\}^{-1}
\end{align}
and given by
\begin{align}
\varphi = \frac{2 \pi n}{\beta} \ (n = 0, \pm 1, \pm 2, \ldots),
\qquad
\chi = 0.
\label{eq:SUSY_vacua}
\end{align}
When the external gauge fields vanish $A_m^B = D^B = 0$, $A_m = D = 0$, then the
vacua preserve all SUSY.
The energy density corresponding to these vacua is exactly zero $\mathcal{E} = 0$.

%%%%%%%
\section{BPS solitons and chiral soliton lattice ground states}
\label{sec:soliton}
In this section, 
we construct BPS domain walls 
and show that 
the ground state is 
a soliton lattice 
consisting of an array of BPS soliton and anti-BPS soliton, 
either in a strong magnetic field 
and/or large chemical potential, 
or in the presence of a background fermion condensate.

%%%%%%%%%%%%%
\subsection{BPS conditions}
We are now in a position to examine the BPS states in the model \eqref{eq:SChPT_superfield2}.
The vanishing conditions on the SUSY transformations of fermions provide
the BPS equations:
\begin{align}
\delta \psi_{\alpha} =& 
%\ i \sqrt{2} (\sigma^m)_{\alpha \dot{\alpha}}
% \bar{\xi}^{\dot{\alpha}} \del_m \varphi + \sqrt{2} \xi_{\alpha} F 
%= 
i \sqrt{2} (\sigma^3)_{\alpha \dot{\alpha}} \bar{\xi}^{\dot{\alpha}} 
\left(
\del_3 \varphi \mp e^{i \eta} F
\right) = 0,
\notag \\
\delta \lambda_{\alpha} =& \ (\sigma^{mn})_{\alpha} {}^{\beta} \xi_{\beta} F_{mn} + i \xi_{\alpha} D = 0,
\notag \\
\delta \lambda^B_{\alpha} =& \ (\sigma^{mn})_{\alpha} {}^{\beta}
 \xi_{\beta} F^B_{mn} + i \xi_{\alpha} D^B = 0.
\label{eq:BPS_eq}
\end{align}
where we have assumed the half SUSY projection on the parameters $\xi_{\alpha} = \mp i e^{i \eta}
(\sigma^3)_{\alpha \dot{\alpha}} \bar{\xi}^{\dot{\alpha}}$.
Here $\eta$ is a phase factor which will be set to zero in the following.
The first condition in Eq.~\eqref{eq:BPS_eq} implies the following BPS equation:
\begin{align}
\varphi' \mp \frac{2m}{\beta} 
%e^{i \eta} 
\sin \left( \frac{\beta}{2}
 \bar{\varphi} \right)
\left\{
\cos
\left(
\frac{\beta}{2} (\varphi - \bar{\varphi})
\right)
+
\frac{\beta^2}{2} i
\sin
\left(
\frac{3 \beta}{2} (\varphi - \bar{\varphi})
\right)
A_m^B \del^m (\varphi - \bar{\varphi})
\right\}^{-1} = 0,
\label{eq:BPS_eq2}
\end{align}
where the prime stands for the differentiation with respect to $x^3 = z$.
It is convenient to decompose the complex scalar field into its real and imaginary
parts $\varphi = \phi + i \chi$.
Then, Eq.~\eqref{eq:BPS_eq2} can be  rewritten as follows:
\begin{align}
&
\phi' \mp \frac{2m}{\beta} \sin \left( \frac{\beta}{2} \phi \right) 
\cosh \left( \frac{\beta}{2} \chi \right)
\Bigg[
\cosh (\beta \chi) - \beta^2 \sinh (3 \beta \chi) A_m^B \del^m \chi
\Bigg]^{-1} = 0,
\notag \\
&
\chi' \mp \frac{2m}{\beta} \sinh \left( \frac{\beta}{2} \chi \right) \cos
 \left( \frac{\beta}{2} \phi \right)
\Bigg[
\cosh (\beta \chi) - \beta^2 \sinh (3 \beta \chi) A_m^B \del^m \chi
\Bigg]^{-1} = 0.
\label{eq:BPS_eq3}
\end{align}
The last two conditions in Eq.~\eqref{eq:BPS_eq} can be trivially solved by
$A_m = D = 0$, $A^B_m = D^B = 0$.
Then, the solution to Eq.~\eqref{eq:BPS_eq3} is given by $\chi = 0$ and $\phi$
satisfying the following equation:
\begin{align}
\phi' \mp \frac{2m}{\beta} \sin \left( \frac{\beta}{2} \phi \right) = 0.
\end{align}
This equation with the upper(lower) sign is nothing but the (anti-)BPS equation in the sine-Gordon model.
The solution to this equation is a single sine-Gordon (anti-)kink:
\begin{align}
\phi (z) = \frac{4}{\beta} \mathrm{arctan} \Big[ e^{\pm m (z-c)} \Big],
\label{eq:single-SG}
\end{align}
where $c$ is the integration constant.
Since we have $\phi \to 0, \pm \frac{2 \pi}{\beta}$ at $z \to \pm \infty$, 
these solutions connect the two adjacent vacua $n = 0$ and $n = \pm 1$ in Eq.~\eqref{eq:SUSY_vacua}.
The corresponding energy can be calculated as 
\begin{align}
E =& \ \int^{\infty}_{-\infty} \! dz \, 
\Bigg[
\phi^{\prime 2} + \frac{4m^2}{\beta^2} \sin^2 \left( \frac{\beta
 \phi}{2} \right)
\Bigg]
\notag \\
=& \ 
\int^{\infty}_{-\infty} \! dz \, 
\Bigg[
\phi' \mp \frac{2m}{\beta} \sin \left( \frac{\beta}{2} \phi \right)
\Bigg]^2
\pm 
\frac{8m}{\beta^2} \int^{\infty}_{-\infty} \! dz \, 
\frac{d}{dz} 
\Bigg\{
\cos \left( \frac{\beta}{2} \phi \right)
\Bigg\}
\notag \\
=& \ \frac{16 m}{\beta^2}.
\end{align}

\subsection{Chiral soliton lattice ground state in background gauge fields}
We next study ground states in the presence of the external
background gauge fields.
The equation of motion for the auxiliary field $D$ is given by
\begin{align}
\cosh (\beta \chi) \, A_m^B \del^m \chi = 0.
\end{align}
A solution to this constraint is found to be $A^B_m = (\mu_{\textrm{B}},0,0,0)$ and $\chi = 0$ 
where the constant $\mu_{\textrm{B}}$ is the baryon chemical potential.
It is evident that this external background breaks SUSY.

Now we look for ground states where the quasi-NG boson $\chi$ remains in the
SUSY vacuum $\chi = 0$. We also assume that the field strength (magnetic field) of the
external $\mathrm{U}(1)_{\text{em}}$ gauge field is constant.
Then, the effective Lagrangian for the neutral pion is 
\begin{align}
\mathcal{L}_{\text{SChPT}} \Big|_{\phi} =& \ 
- \del_i \phi \del_i \phi
- \frac{4 m^2}{\beta^2} \sin^2 \left( \frac{\beta}{2} \phi \right)
+ \mu_{\textrm{B}} \beta B_i \del_i \phi,  
\label{eq:SG_magnetic_bg}
\end{align}
where the sum over the 
index $i$ is assumed.
This is nothing but the sine-Gordon model in the external magnetic
field, or the chiral sine-Gordon model. 

Let us assume $B_i =(0,0,B_3)$ 
and a one-dimensional dependence of the configurations.
Then, the energy density is 
\begin{align}
\mathcal{E} =& \ 
(\phi')^2 + \frac{4m^2}{\beta^2} \sin^2 \left( \frac{\beta}{2} \phi
 \right) - \mu_{\textrm{B}} \beta B_3 \phi'.
\end{align}
The last term which is a total derivative term is induced by the external background. 
The equation of motion for $\phi$ is 
\begin{align}
\phi'' = \frac{m^2}{\beta} \sin (\beta \phi).
\end{align}
Note that the last term in Eq.~\eqref{eq:SG_magnetic_bg} does not contribute
to the equation of motion.

A single sine-Gordon (anti-)kink solution 
in Eq.~(\ref{eq:single-SG}) 
remains a solution. 
In this case, the tension
(energy per unit area)  
is 
\begin{align}
E =& \ \frac{16m}{\beta^2} \mp 4 \pi \mu_{\textrm{B}} B_3,
\end{align}
where $\mp$ corresponds to the sign of the (anti-)BPS kink. 
Thus, the second term contributes 
negatively (positively) to 
the tension 
for the (anti-)kink.
Therefore for 
\begin{equation}
   \mu_{\textrm{B}} |B_3| \geq \frac{4m}{\pi  \beta^2},
    \label{eq:susy-CSL}
\end{equation}
equivalent to Eq.~(\ref{eq:critical}) in this parameterization, 
the energy of the BPS solutions 
is negative (zero) and is thus less than (or equal to) that of the SUSY vacua\footnote{
It is known that when there are negative energy states in a SUSY theory, 
there appear ghosts in the theory. See for example
 \cite{BjarkeGudnason:2018aij}.
We expect that these ghosts would be removed by
the dynamics of external fields or other appropriate mechanics discussed elsewhere.
}.
We note that the external backgrounds introduced here do not satisfy
 the last two equations in Eq.~\eqref{eq:BPS_eq} and hence they completely break the 
 remaining SUSY preserved by the BPS sine-Gordon kink.

More precisely, at the boundary of Eq.~(\ref{eq:susy-CSL}), a single kink has a zero tension and is a ground state 
degenerate with the SUSY vacua.
When the inequality in Eq.~(\ref{eq:susy-CSL}) holds, 
the ground state is more general solution, 
given by the CSL
\begin{align}
\phi (z) = \pm \frac{2}{\beta} \mathrm{am} \Big( m k^{-1} (z-c),k \Big)
\end{align}
where $\mathrm{am} (x,k)$ is the amplitude of the Jacobi elliptic
function, $c$ is the integration constant and $0 \le k \le 1$ is the elliptic modulus parameter, 
determined below.

The energy per unit volume is given by
\begin{align}
E =& \frac{m}{2k K(k)}
\Bigg[
\frac{4m}{\beta^2} 
\Big\{
\frac{2 E(k)}{k} + \left( k - \frac{1}{k} \right) K(k)
\Big\}
\mp 2 \pi \mu_{\textrm{B}} |B_3|
\Bigg], 
\label{eq:csl_energy}
\end{align}
where $K(k)$ and $E(k)$ are the complete elliptic integral of
the first and the second kinds, respectively.
Again, we take the upper sign for the ground state.
The extremization of 
 the energy in 
the expression \eqref{eq:csl_energy} gives the
condition for $k$:
\begin{align}
\frac{E(k)}{k} = \frac{\pi \mu_{\textrm{B}} \beta^2}{4m} |B_3|.
\end{align}
Then, this condition actually give 
the minimum energy, given by
\begin{align}
E = \frac{4m}{\beta^2} \left( k - \frac{1}{k} \right) K(k) < 0.
\end{align}
This is clearly less than (or equal to) the energy of the SUSY vacua.
As a larger background $\mu_{\textrm{B}} |B_3|$, the elliptic modulus $k$ is smaller and the lattice spacing becomes shorter 
($k \to 0$ as 
$\mu_{\textrm{B}} |B_3| \to \infty$). 
At the boundary of the inequality 
(\ref{eq:susy-CSL}), 
the elliptic modulus is $k=1$  recovering 
a single soliton with $E=0$.

\subsection{Chiral soliton lattice ground state 
in fermion condensates}
We next examine the effects of fermion contributions in the background vector multiplets.
When we keep the fermions in the background multiplets $V$ and $V^B$ and
drop the chiral fermion $\psi$ in $\Phi$, we find that the interaction
Lagrangian gives
\begin{align}
&
- \frac{\beta}{2} \int \! d^4 \theta \, 
\cos \left[ \frac{\beta}{2} (\Phi - \Phi^{\dagger}) \right]
V^B W_{\alpha} D^{\alpha} \Phi + \text{h.c.}
\notag \\
\sim
& \ 
\beta \cos (\beta \chi)
\Bigg\{
- A_m^B D \del^m \chi + A_m^B F^{mn} \del_n \chi 
+ \frac{1}{2} \varepsilon^{mnpq} A_p^B F_{mn} \del_q \phi
\notag \\
& \qquad \qquad \qquad
+ 
\frac{i}{2}
\Big(
\lambda^B \bar{\sigma}^m \lambda - \bar{\lambda} \bar{\sigma}^m \lambda^B
\Big) \del_m \phi
-
\frac{1}{2}
\Big(
\lambda^B \bar{\sigma}^m \lambda + \bar{\lambda} \bar{\sigma}^m \lambda^B
\Big)
\del_m \chi
\notag \\
& \qquad \qquad \qquad
+
\frac{1}{2} (\lambda^B \lambda) F
+
\frac{1}{2} (\bar{\lambda} \bar{\lambda}^B) F^{\dagger}
\Bigg\}.
\end{align}
Assuming that the background fermion bilinear consisting of 
the gaugino $\lambda$ and the superpartner  $\lambda^B$  of 
the baryon gauge field 
\begin{equation}
  j^m = \frac{i}{2}(\lambda^B \bar{\sigma}^m \lambda - \bar{\lambda} \bar{\sigma}^m \lambda^B)  \label{eq:fermion-cond}
\end{equation}
gets a constant VEV $\langle j^i \rangle$, then the energy density for the static
field $\phi$ is given by
\begin{align}
\mathcal{E}|_{\phi} =& \ 
(\del_i \phi)^2 + \frac{4m^2}{\beta^2} \sin^2 \left( \frac{\beta}{2}
 \phi \right)
\mp  \beta \langle j^i \rangle \del_i \phi.
\end{align}
Since the third term is the total derivative, it never contributes to the equation of motion for $\phi$.
However, it does contribute to the energy.
For example, for a single sine-Gordon (anti-)kink solution, we find
its tension as
\begin{align}
E = \frac{16m}{\beta^2} \mp 4 \pi \langle j^3 \rangle
\end{align}
where we have assumed 
$\langle j^i \rangle = \delta^{i3} \langle j^3 \rangle$ 
\footnote{
It is expected that this specific orientation of the fermion bilinear is determined by
 quantum effects on a background oriented in a certain direction.
 }.
The second term with upper (lower) 
 sign apparently provides the negative contribution for the 
positive (negative) VEV. 
Therefore, when the inequality 
\begin{equation}
\left| \frac{i}{2}(\lambda^B \bar{\sigma}^m \lambda - \bar{\lambda} \bar{\sigma}^m \lambda^B) \right|
\geq  \frac{4m}{\pi \beta^2}  
\label{eq:current_bound}
\end{equation}
holds, the CSL is the ground state in the presence of
this fermion condensation.
This is the SUSY counterpart of the baryon density and
external magnetic field.
We note that the fermionic background generically breaks 
all the SUSY preserved by the sine-Gordon kink, irrespective of the orientations of
 the fermion bilinear.

It is useful to consider the scale of the fermion condensation 
(\ref{eq:fermion-cond}).
A famous example is the gaugino condensation for which a fermion bilinear acquires a non-zero VEV due to the instanton effects \cite{Davies:1999uw}. In this case, the fermion bilinear should behave as $\Lambda^3$ by the dimensional analysis. Here $\Lambda$ is the dynamical scale of the model.
Other examples in which fermion bilinear develops a VEV can be found in studies of the early universe.
For example, some kinds of dynamics have been studied to generate a VEV of fermionic current density in the locally AdS spacetime \cite{Bellucci:2018rzd, Bellucci:2019viw}.
It has been discussed that the temporal component of a fermion current dominates the energy density for the gauge field in the early (de Sitter) universe and it causes the cosmic inflation \cite{Alexander:2011hz, Alexander:2014uza}.
In any event, it is therefore natural that the left-hand side in Eq.~\eqref{eq:current_bound} behaves like $\Lambda^3$ and we expect that the CSL ground state happens at $\Lambda \gg \sqrt[3]{\frac{m}{\beta^2}}$.

%%%%%%%%%%%%%%%%%%%
\section{Summary and Discussion}
\label{sec:summary}

In this paper, we have constructed 
the neutral pion sector of   
a manifestly supersymmetric  
extension of the $\SU(2)$ 
chiral Lagrangian with the WZW term 
in magnetic field at finite density, 
which is the SUSY chiral sine-Gordon model.
We then 
have constructed (anti-)BPS domain wall 
(chiral soliton) 
and shown that 
the tension of the domain wall becomes negative and 
the ground state is CSL 
where SUSY is broken, 
when the background magnetic field 
and/or chemical potential are large enough  
or when there is a background fermion 
condensates 
in the form of a fermion bilinear consisting of 
the superpartner of the baryon gauge field and the gaugino.

In this paper, we have been able to construct 
only the neutral pion sector of the SUSY chiral Lagrangian 
with the WZW term in the superfield formalism.
Constructing the full Lagrangian including charged pions 
remains a future problem. 
Once the charged pions can be taken into account, 
one can also discuss 
domain-wall Skyrmions as were studied in QCD 
\cite{Eto:2023lyo,Eto:2023wul}.
The domain-wall Skyrmions  
are composite states of a domain wall and Skyrmions, initially introduced in the  field theoretical models in 3+1 dimensions~\cite{Nitta:2012wi,Nitta:2012rq,Gudnason:2014nba,Gudnason:2014hsa,Eto:2015uqa,Nitta:2022ahj} and in 2+1 dimensions~\cite{Nitta:2012xq,Kobayashi:2013ju,Jennings:2013aea}. 
\if0 
In condensed matter physics, the 2+1 dimensional variant has been theoretically investigated~\cite{PhysRevB.99.184412,KBRBSK,Ross:2022vsa,Amari:2023gqv,Amari:2023bmx,PhysRevB.102.094402,Kim:2017lsi,Lee:2022rxi} and experimentally observed in chiral magnets~\cite{Nagase:2020imn,Yang:2021}, 
\fi
%in which case
If 
a baby Skyrmion in 2+1 dimensions, 
supported by $\pi_2(S^2) \simeq {\mathbb Z}$, 
is absorbed into a domain wall, 
it becomes a sine-Gordon soliton supported by 
$\pi_1(S^1) \simeq {\mathbb Z}$ in the domain-wall world line.
Similarly, 
a Skyrmion in 3+1 dimensions, supported by $\pi_3(S^3) \simeq {\mathbb Z}$, 
is absorbed into a chiral soliton 
to become 
a topological lump (or baby Skyrmion) 
supported by $\pi_2(S^2)\simeq {\mathbb Z}$ in 
the solitons's worldvolume \cite{Eto:2023lyo}. 
SUSY extensions of these cases are worth to be investigated.\footnote{
However, it is worth to mention that 
a supersymmetric extension of the Skyrme model 
does not contain quadratic derivative term
\cite{Gudnason:2015ryh}, 
allowing only non-BPS solitons \cite{Gudnason:2016iex}.
} 
In particular, whether such a composite state can be a $1/4$ BPS state \cite{Tong:2005un,Eto:2006pg,Shifman:2007ce,Shifman:2009zz,Isozumi:2004vg,Eto:2004rz,Eto:2005sw} is one of interesting direction to be explored.

In dense QCD, 
a CSL state is unstable 
due to the charged pion condensation 
in a region of higher density and/or stronger magnetic field, 
asymptotically expressed 
at $B$ larger than $
    {16 \pi^4 f_{\pi}^4}/{\mu_{\rm B}^2}$ \cite{Brauner:2016pko}  
above which tachyon appears and 
the charged pions are condensed.
Consequently, the CSL becomes unstable, 
where 
 an Abrikosov's vortex lattice was proposed as a consequence of the charged pion condensation~\cite{Evans:2022hwr,Evans:2023hms}.
It is an open question whether 
such an instability is present  
in our SUSY extension. 
%Although SUSY is completely broken in 
%the total configuration of CSL, 
%each (anti-)soliton 
%is an (anti-)BPS state 
%preserving a half SUSY and 
%is stable at least when 
%individual solitons are well separated.
%We thus expect that the stability is enhanced 
%in our SUSY extension. 
Another characteristic of SUSY theory is the following.
At least when the solitons are well separated, 
individual soliton is either BPS or anti-BPS, 
thereby approximately maintaining BPS saturation 
({\it i.e.} energy = central charge). 
This leads us to expect the stability of each soliton against quantum corrections.
As the density of solitons increases and the 
inter-soliton distance decreases, 
the interactions between BPS and anti-BPS solitons become significant. 

We have discussed 
bosonic  and fermionic backgrounds  
separately. 
If there are both kinds of backgrounds 
simultaneously in different spatial directions, 
say $B_i =(0,0,B_3)$ 
and $\langle j^i \rangle = (0,j^2,0)$, 
sine-Gordon solitons have two preferable directions. 
It is an open question whether 
the ground state is still a CSL in 
a certain direction  linearly determined from the two directions or 
a soliton junction.

In this paper, we have discussed only 
the leading order ${\cal O}(p^2)$ of the 
ChPT.
The higher order term of SUSY ChPT 
was discussed in Ref.~\cite{Nitta:2014fca}
by using SUSY higher derivative 
terms free from ghost and auxiliary field problem 
\cite{Khoury:2010gb,Khoury:2011da,Koehn:2012ar,Adam:2011hj,Farakos:2012qu,Nitta:2014pwa,Nitta:2014fca,Nitta:2015uba,Nitta:2020gam}. 
SUSY may help us to consider ChPT  
in a more controllable manner 
and may eventually uncover 
a phase structure of 
non-SUSY QCD under extreme conditions such as 
strong magnetic field and/or rapid rotation.

As mentioned in the introduction, 
chiral magnets also admit solitonic ground states 
such as CSL and a Skyrmion lattice.
This model can be written as 
a ${\mathbb C}P^1$ model 
with a background non-Abelian gauge field \cite{Schroers:2019hhe}.
It can be made supersymmetric 
and can be embedded into string theory \cite{Amari:2023gqv}. 
However, the bosonic part of the model was given only  in terms of  component fields 
and the fermionic part is not given \cite{Amari:2023gqv}, and so 
it is a future problem to construct the chiral magnets in
the superfield formalism to study e.~g. SUSY breaking. 
It is desirable to have a unified understanding of 
inhomogeneous ground states in 
SUSY field theories 
with background gauge fields.

\begin{acknowledgments}
This work is supported in part by 
 JSPS KAKENHI [Grants No.~JP22H01221 (MN)] and the WPI program ``Sustainability with Knotted Chiral Meta Matter (SKCM$^2$)'' at Hiroshima University (MN).
\end{acknowledgments}

%%%%%%%%%%%%%%%%%%%%%
\begin{appendix}
\section{Fayet-Iliopoulos term}
\label{sec:FI}
In this appendix, we show that the introduction of the Fayet-Iliopoulos term in our model causes inconsistencies.
The Lagrangian may be given by
\begin{align}
\mathcal{L} = \mathcal{L}_{\text{SChPT}} + \xi \int \! d^4 \theta \, V,
\end{align}
where $\xi$ is the Fayet-Iliopoulos parameter.
Then, the $D$-term condition is found to be
\begin{align}
- \beta \cosh (\beta \chi) A^B_m \del^m \chi + \frac{1}{2} \xi = 0.
\end{align}
For the given background $A^B_m = (\mu_{\textrm{B}},0,0,0)$, we have the condition
\begin{align}
\beta \mu_{\textrm{B}} \cosh (\beta \chi) \dot{\chi} = - \frac{1}{2} \xi.
\end{align}
The solution to this equation is 
\begin{align}
\chi (t) = \frac{1}{4} \mathrm{arctan} \Big[ - \frac{\xi t}{2 \mu_{\textrm{B}}} t + c \Big],
\label{eq:FI_solution}
\end{align}
where $c$ is the integration constant.
On the other hand, for $\phi = 0$, the equation of motion for $\chi$ is given by
\begin{align}
&
- \beta \sinh(\beta \chi) \del_m \chi \del^m \chi
- \frac{2m^2}{\beta} \frac{\sinh (\beta \chi)}{\cosh^2 (\beta \chi)}
+ \beta \sinh (\beta \chi) A_m^B F^{mn} \del_n \chi
\notag \\
&
- \del_m
\Bigg[
- 2 \cosh (\beta \chi) \del^m \chi
+ \cosh (\beta \chi) A_m^B F^{mn} \del_n \chi
\Bigg] = 0.
\end{align}
Since the solution \eqref{eq:FI_solution} explicitly depends on $\xi$ but the equation of motion does not, 
it is obvious that the solution \eqref{eq:FI_solution} never satisfies the equation of motion unless $\xi = c = 0$ and $\chi=0$.
\end{appendix}

%%%%%%%%%%%%%%%%%%%%%%%%%%%
\bibliographystyle{jhep}
\bibliography{reference.bib}

%\begin{thebibliography}{99}
%\end{thebibliography}

\end{document}